\documentclass[preprints,article,accept,moreauthors]{mdpi}
\firstpage{1} 
\pubvolume{1}
\issuenum{1}
\articlenumber{0}
\pubyear{2022}
\copyrightyear{2022}
\datereceived{} 
\dateaccepted{} 
\datepublished{} 
\hreflink{https://doi.org/} % If needed use \linebreak
\Title{Applications of the source-frequency phase-referencing technique for ngEHT observations}

\TitleCitation{Applications of the source-frequency phase-referencing technique for ngEHT observations}

\newcommand\sgra{Sgr\,A$^{*}$\,}
\newcommand\aap{A\&A}
\newcommand\nat{Nature}
\newcommand\pasj{PASJ}
\newcommand\apjl{ApJL}
\newcommand\apj{ApJ}
\newcommand\apjs{ApJS}
\newcommand\aj{AJ}
\newcommand\aapr{AApR}
\newcommand\araa{ARAA}
\newcommand\mnras{MNRAS}
\newcommand\pasp{PASP}
\Author{Wu Jiang $^{1,2,}$*\orcidA{}, Guang-Yao Zhao $^{3,}$*\orcidB{}, Zhi-Qiang Shen $^{1,2,}$\orcidC{}, Mar\'{\i}a Rioja$^{4,5,6.}$\orcidD{}, Richard Dodson$^{4,}$\orcidE{}, Ilje Cho$^{3,}$\orcidF{}, Shan-Shan Zhao$^{1,2,}$\orcidG{}, Marshall Eubanks$^{7,}$\orcidH{} and Ru-Sen Lu$^{1,2,8,}$\orcidI{}}

\AuthorCitation{Jiang, W. et al.}
\address{%
$^{1}$ \quad Shanghai Astronomical Observatory, 
 Chinese Academy of Sciences, Shanghai 200030, China.\\
$^{2}$ \quad Key Laboratory of Radio Astronomy,  
 Chinese Academy of Sciences, Nanjing 210008, China.\\
$^{3}$ \quad Instituto de Astrof\'{\i}sica de Andaluc\'{\i}a-CSIC, Glorieta de la Astronom\'{\i}a s/n, 18008 Granada, Spain.\\
$^4$ \quad {ICRAR, M468, The University of Western Australia, 35 Stirling Hwy, Crawley, Western Australia, 6009, Australia}\\
$^5$ \quad {CSIRO Astronomy and Space Science, PO Box 1130, Bentley WA 6102, Australia}\\
$^6$ \quad{Observatorio Astron\'omico Nacional (IGN), Alfonso XII, 3 y 5, 28014 Madrid, Spain}\\
$^{7}$ \quad {Space Initiatives Inc, Newport, VA 24128, USA}\\
$^{8}$ \quad Max-Planck-Institut f\"ur Radioastronomie, Auf dem H\"ugel 69, D-53121 Bonn, Germany.
}

\corres{Correspondence: jiangwu@shao.ac.cn; gyzhao@iaa.es}
\abstract{The source-frequency phase-referencing (SFPR) technique has been demonstrated to have great advantages for mm-VLBI observations. By implementing simultaneous multi-frequency receiving systems on the next generation Event Horizon Telescope (ngEHT) antennas, it is feasible to carry out a frequency phase transfer (FPT) which could calibrate the non-dispersive propagation errors and significantly increase the phase coherence in the visibility data. Such increase offers an efficient approach for weak source or structure detection. SFPR also makes it possible for high precision astrometry, including the core-shift measurements  up to sub-mm wavelengths for Sgr\,A* and M\,87* etc. We also briefly discuss the technical and scheduling considerations for future SFPR observations with the ngEHT.} 

\begin{document}

%%%%%%%%%%%%%%%%%%%%%%%%%%%%%%%%%%%%%%%%%%
\section{Introduction}

The Very Long Baseline Interforemetry (VLBI) technology can achieve the highest spatial angular resolution by linking intercontinental telescopes to form a virtual telescope, whose aperture size is equal to the longest baseline in the array.  However, the wavefront arriving at each telescope suffers from various phase fluctuations when propagating through the atmosphere. This is even more severe at the millimeter and sub-millimeter (sub-mm) wavelengths as the phase dispersion is in proportion to the observing frequency. A novel technique called frequency phase transfer (FPT) \citep{2005A&A...433..897M} or source-frequency phase-referencing (SFPR) \citep{2011AJ....141..114R} is proposed to mitigate the fast phase fluctuations at the shorter wavelengths, by referring to the phases at the longer wavelength observed close in time. The phases could be purified by two step calibrations. The first step is the FPT calibration, where the non-dispersive phase errors, such as the tropospheric phase errors and the geometric antenna position errors, are removed. Furthermore, the unmodeled ionospheric delay and the instrumental phase offsets between the two wavelengths can be further eliminated by observations of a nearby calibrator. After the SFPR calibrations, the remaining phases just reflect the true high frequency visibilities and the frequency-dependent shift in the positions, e.g., the frequency-dependent location of the jet cores (core-shift) \citep{1998AA...330..79L}. SFPR could also help to reliably align the molecular line emission seen at different frequency bands (e.g., \citep{2018NatCo...9.2534Y}). It is of great advantages in probing weak sources and high precision astrometric measuring for the (sub-)mm-VLBI. \\
The capability of simultaneously receiving at four frequency bands (K/Q/W/D) makes the Korea VLBI Network (KVN) a unique prototype and instrument for the FPT/SFPR observations \citep{2015AJ....150..202R, 2019JKAS...52...23Z}. The capability of fast-switching among receivers at the Very Long Baseline Array (VLBA) also makes it possible to carry out FPT/SFPR observations up to the 3\,mm band \citep{2011AJ....141..114R, 2018APJL...853..14J}, although the switching cycle time introduces coherence losses \citep[see Fig. 6 in][] {2020AARv..28....6R}.
\begin{itemize}
    \item{ ngEHT and the necessary of SFPR }
    
    Based on the success of capturing the first images of two nearby supermassive black holes with the original Event Horizon Telescope (EHT), one at the center of the distant Messier 87 galaxy (M87*) \citep{2019ApJ...875L...1E} and the other at our Milky Way galaxy center (Sgr\,A*) \citep{2022ApJ...930L..12E}, the next generation Event Horizon Telescope (ngEHT) will expand the existing array (new sites) \citep{2021ApJS..253....5R} and upgrade the technological deployments (receiving capabilities) significantly \citep{2019BAAS...51g.256D}. It aims to sharpen our view of the black holes and address fundamental questions about the accretion and jet launching process, together with more black hole shadows captured and even making black hole "movies".
%\latex\ \footnote{\url{http://www.latex-project.org/}} is a document markup

Although the sensitivity of ngEHT would be greatly improved with an ultra-wide bandwidth, the baseline sensitivity will still be limited due to the short coherent integration time at sub-mm wavelengths (a typical coherence time is $\sim$10 seconds at 230\,GHz \citep{2019ApJ...882...23B,2019ApJ...875L...3E} and even shorter at 345\,GHz) and the small dish size of most antennas. SFPR can overcome the coherence time limitation at sub-mm wavelengths. 
As demonstrated in a separate technical paper of this special issue, the coherence time of the high frequency by referring to the low frequency band could be increased more than 100 folds and extended to hour(s) in the simulations. See Rioja et al. in the same issue for more details. The detection threshold relies on the lower frequency rather than the higher one. Using a typical value of 10-15\,sec at 85\,GHz, the flux density threshold for targets would become one magnitude lower ($\sim10$\,mJy) and the number of targets would be hundreds under the array sensitivity. We have estimated the SFPR errors that would be introduced when referencing the 255 or 340\,GHz data to 85\,GHz, with an angular separation of 10$^{\circ}$ between sources. With simultaneous multi-frequency observations and intra-source switching times between zero and 10 minutes, the astrometric precision is about 3\,{\textmu}as and dominated by the static ionospheric residuals. These would make ngEHT more powerful for both astrophysical and astrometric applications.   

%This enables many interesting scientific possibilities.
\end{itemize}

%%%%%%%%%%%%%%%%%%%%%%%%%%%%%%%%%%%%%%%%%%
\section{Scientific applications}
\subsection{Sgr\,A* and M\,87* }
Sgr A* and M\,87* are the prime targets for demonstrating the  application of SFPR to observational studies of black holes and jets. SFPR can help reduce the phase error budgets from the atmosphere and instruments, while increasing the coherence time, and thus improve the dynamical range of imaging. Furthermore, SFPR will provide precise measurements to understand the event-horizon-scale structure adjacent to the supermassive black holes. 
\begin{itemize}
    \item Possible core-shift detection of Sgr A* % [IC \& GYZ]
     %\rd{recall that for resolved sources you will need a hub-and-spoke array configuration. Maybe not for SgrA* -it is so strong.}
     %\gy{I think that is the whole idea of the ngEHT, adding small dishes around the 'anchor' stations to fill the uv-plane.}
     
     The mm/sub-mm radio emission from \sgra can be produced by two generic models: an accretion flow itself \citep{narayan_1995,yuan_2003} and/or an outflow \citep{falcke_2000}. To discriminate the dominant emission models of \sgra, the core shift \citep[e.g.,][]{blandford_1979, lobanov_1998} can be used without resolving its structure. 
     %
     %In VLBI observations, an optically thick surface (the photosphere) is shown as a radio core at a given observing frequency and moves toward the central SMBH with increasing frequency. The core shift can give a hint of not only the jet direction even when a clear structure cannot be resolved in sub-pc scale, but also a precise position of the central SMBH at upstream end of a jet. On the other hand, and especially for \sgra, the core shift can also be shown by the asymmetric structure of an accretion flow which has been predicted as the Doppler-boosted side of the rotating accretion disk. However the expected amount of the core shifts is different in each scenario so that they can be discriminated by measuring it precisely (Fraga-Encinas et al.~in prep.).
     %
     As for the jet model, based on GRMHD simulations, \citet{moscibrodzka_2014} suggested the core shift of $\sim130$\,{\textmu}as at 22-43\,GHz and $\sim60$\,{\textmu}as at 86-230\,GHz. In a recent study by Fraga-Encinas et al. (in prep.), the core shift of \sgra is predicted from both accretion disk and jet model with different inclination angles. According to their results, a clear difference in core shift between the two scenarios is shown. Especially at small inclination angle, as has been suggested in recent studies \citep{gravity_2018b, issaoun_2019, cho_2022}, the expected core shift at 22-43\,GHz is $\lesssim10$\,{\textmu}as in accretion disk model while $\gtrsim100$\,{\textmu}as in jet model. 
     Our preliminary core shift measurements with the Korean VLBI Network (KVN) and the Very Long Baseline Array (VLBA) at the same frequencies show $\sim$100\,{\textmu}as (I.~Cho et al. in prep). However the robustness has been relatively less due to large astrometric uncertainties which are mainly originated from 1) the large beam size (for KVN) and 2) the frequency switching mode (for VLBA). Each difficulty can be perfectly overcome through the ngEHT with the dual/triple band receiving capability. 
     %%%
     %%%
    \item Connecting the jet and the Black hole for M87* % [GYZ]
    % \gy{Whether we keep this item would highly depend on the ongoing GMVA 3 mm analysis results. If we can already answer how the BH and jet are co-located with a single image, then we don't need MF observations.}
    % \gy{(PLACEHOLDER) 
    
    The EHT 2017 image of M87* has revealed the shadow of the central SMBH \citep{2019ApJ...875L...1E}. The EHT observations, however, were unable to reliably detect and image the inner jet, likely due to sensitivity limitations and the lack of short baselines in the UV-coverage.
    At longer wavelengths, we see a well-collimated jet but the emission is optically-thick and we are only able to see the $\tau$=1 surface and the downstream optically-thin jet\citep[][]{2011Natur...477..185H}. %(but see also %\gy{Lu et al. submitted} for the recent 3.5\,mm results). 
    Furthermore, the resolution at longer wavelengths are not enough to resolve the shadow\citep{2021ApJ...911L..11E}.
    It remains uncertain how exactly the SMBH and the jet are connected.
    The ngEHT will improve the dynamic range of the 1.3\,mm images which could enable the detection of the extended jet emission. However, it could be still challenging due to the steep spectrum of the jet. 
    SFPR covering 86-345 GHz bands offers an alternative way to reliably determine the relative location of the SMBH we see at 1.3\,mm and the jet core at longer wavelengths.
    This is critical in understanding how black holes launch powerful, collimated jets\cite[e.g.][]{blandford19}.
    % }
    %
    %\item proper motion of Sgr A*
    %\rd{recall that for PM you will need conventional PR -- which is possible with 4 antennas per site.}
    %\gy{Is it technically easier to implement some dual-beam system? Otherwise, we should probably leave the proper motion to lower frequency arrays.}
    %\wj{How about two antennas per site? as we can alternatively switch from target-calibrator A to target-calibrator B or C. This could be doable with the small dishes around the 'anchor' stations as GY mentioned. Then how close the two antenna should be?}
    %\rd{On consideration we think the purpose of this document should be to set requirements, not design the solution. For proper motion studies conventional PR is required. What would science would proper motion studies provide?}
    %Current proper motions of SgrA* still surfer from the scattering as measured at 43G \citep{2004ApJ...616..872R}, if can go to higher frequency this effect should be reduced. Eventually it can obtain a high precision of the parallax measurement, which gives the distance by VLBI. This would be a good start to other targets as well (section 2.3). \citep{2011ApJ...735...57B}, absolute astrometry needs cluster/paired antennas in each site. reveal the nature of the ISM scattering of Sgr~A* 
    %\ssz{The scattering effect is frequency-dependent, we can deduce the scattering phase screen properties from the multi-frequency observation.}
    %\gy{Is SFPR required for this topic?}
    %\wj{move the scattering effect of Sgr~A* to be included in the proper motion item}
    %
\end{itemize}

\begin{figure}[H]
\includegraphics[width=14 cm]{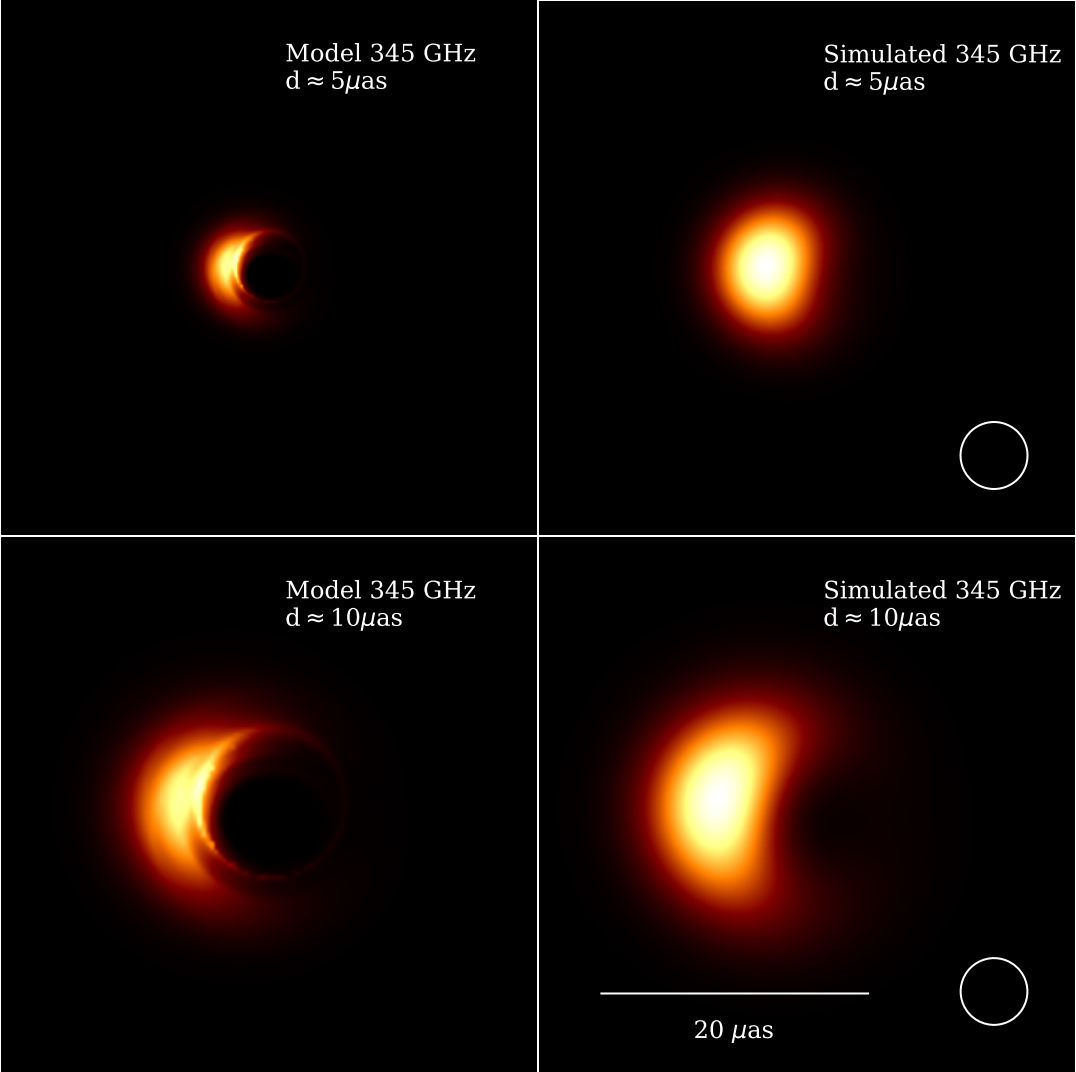}
\caption{Model images of M84 with two different black hole masses. The images (right column) are reconstructed based on simulated ngEHT observations at 345\,GHz. The empty white ring at the right bottom corner of the right panel plots is the synthesized beam of ngEHT at 345\,GHz.\label{fig1}}
\end{figure}  

\subsection{ Detection of weak sources and structures }
 \begin{itemize}
  \item Toward more supermassive black hole shadows
  
  %[table/ mass /shadows size /flux density/distance (EHTC/EAhi/duplicated with other WPs? high priority list)]
  With the increased coherent integration time, black holes whose radio emissions are weak but shadow sizes are relatively large can be detected by ngEHT. According to the prediction of a semi-analytic spectral energy distribution model \citep{2009ApJ...706..960H}, there should be a dozen of additional sources with their horizon-scale structure resolved the ngEHT observing at 345 GHz \citep{2021ApJ...923..260P}. M\,84, M\,104, and IC\,1459 are the prominent candidates on the priority list. These targets have a correlated flux density of several tens mJy \citep{2021APJL...922..16J} and a shadow size of $\sim$10 \,{\textmu}as. The sources could be directly fringed with a short solution interval and a relatively high signal-to-noise ratio at 85\,GHz, that guarantees the quality of phases to be transferred to higher frequencies. The predicted sizes of black hole shadows are comparable to the resolution achievable by ngEHT at 345\,GHz. It provides further test samples of black holes whether or not described by the Kerr metric besides M87* and Sgr A*. Vice versa, combining the diameter measurements of black hole shadows with GRMHD simulations, plus an independent distance measurement, can be used to determine physical parameters of black holes (e.g., mass, orientation, and spin etc). 

Towards understanding of black holes, we are still on the road of pursuing precise measurements and conclusive evidences. In the case of M84 (z = 0.00339, D = 18.4\,Mpc), the mass of the central super massive black hole is 8.5$\times10^8\,M_\odot$ measured by the gas kinematics \citep{2010APJ...721..762W}, or 1.8$\times10^9\,M_\odot$ estimated from velocity dispersion \citep{2004AJ....127..119L}. Therefore, the diameter $d$ of the black hole shadow would be about 5\,{\textmu}as or 10\,{\textmu}as, respectively. M\,84 has a correlated flux of about 80 mJy at 86\,GHz (Wang et al. in press), while the baseline sensitivity of ngEHT at 86\,GHz would achieve several mJy, that would guarantee the phase solutions with a signal-to-noise ratio high enough  to be transferred to 345\,GHz. As shown in Fig. 1, the black hole mass could be independently constrained by the angular size of the shadow. It also indicates that ngEHT with SFPR could image a batch of black hole shadows whose diameters are $\sim$10\,{\textmu}as. SFPR could increase the coherent integration time that promises a firm fringe detection at 345\,GHz, as well as  high dynamic range imaging with sub-diffraction limited resolution \citep{2017ApJ...838....1A}. 
%\begin{itemize}
%\end{itemize}
%\unskip
%\begin{equation}
%T_{b, app} = 1.22\times10^{12} %\frac{S_{core}(1+z)}{\nu^2\theta_{core}^2} \,K,
%\end{equation}
\item Detection of cosmic sources at 1mm

Based on the radio luminosity function, the number of AGNs detectable at millimeter is almost inversely proportional to the array sensitivity. Besides of detecting the horizon structure of faint nearby SMBHs, SFPR could be used to increase the detection of cosmic sources at short wavelengths. The flux threshold of SFPR detection will be $\sim10$\,mJy through simulations. According to the ALMA calibrator catalog \footnote{\url{https://almascience.eso.org/sc/}}, there would be more than nine hundred sources observable. These sources have a correlated flux (considering a resolving factor of $\sim$0.16 with a baseline length of 5000\,km) higher than 10\,mJy and a flat spectrum from 85\,GHz to 345\,GHz. With the increased sensitivity of ngEHT which is further enhanced by SFPR, that provides more diverse samples approachable at the upstream of jets for physical parameter statistics, such as the brightness temperature of the mm-core and the collimation profile of the jet base \citep{2016ApJ...833...56A,2021NatAs...5.1017J}, as well as sub-structures in the core region \citep{2018NatAs...2..472G}.
\end{itemize}
\subsection{Micro-arcsecond astrometry to the black holes }
  SFPR enables the VLBI astrometry at millimeter/sub-millimeter wavelengths with a precision of several {\textmu}as. That means 0.01\,pc motions of targets can be measured within a distance of Gpc. By source-frequency phase referencing, the location of black hole could be pinpointed  \citep{2021APJL...922..16J}. It enables the micro-arcsecond astrometry to the black hole itself in the ngEHT era. % \gy{This could be possible with SFPR too, and this paper is focused on SFPR, not MFPR}.
  
\begin{figure}[H]
\includegraphics[width=14 cm]{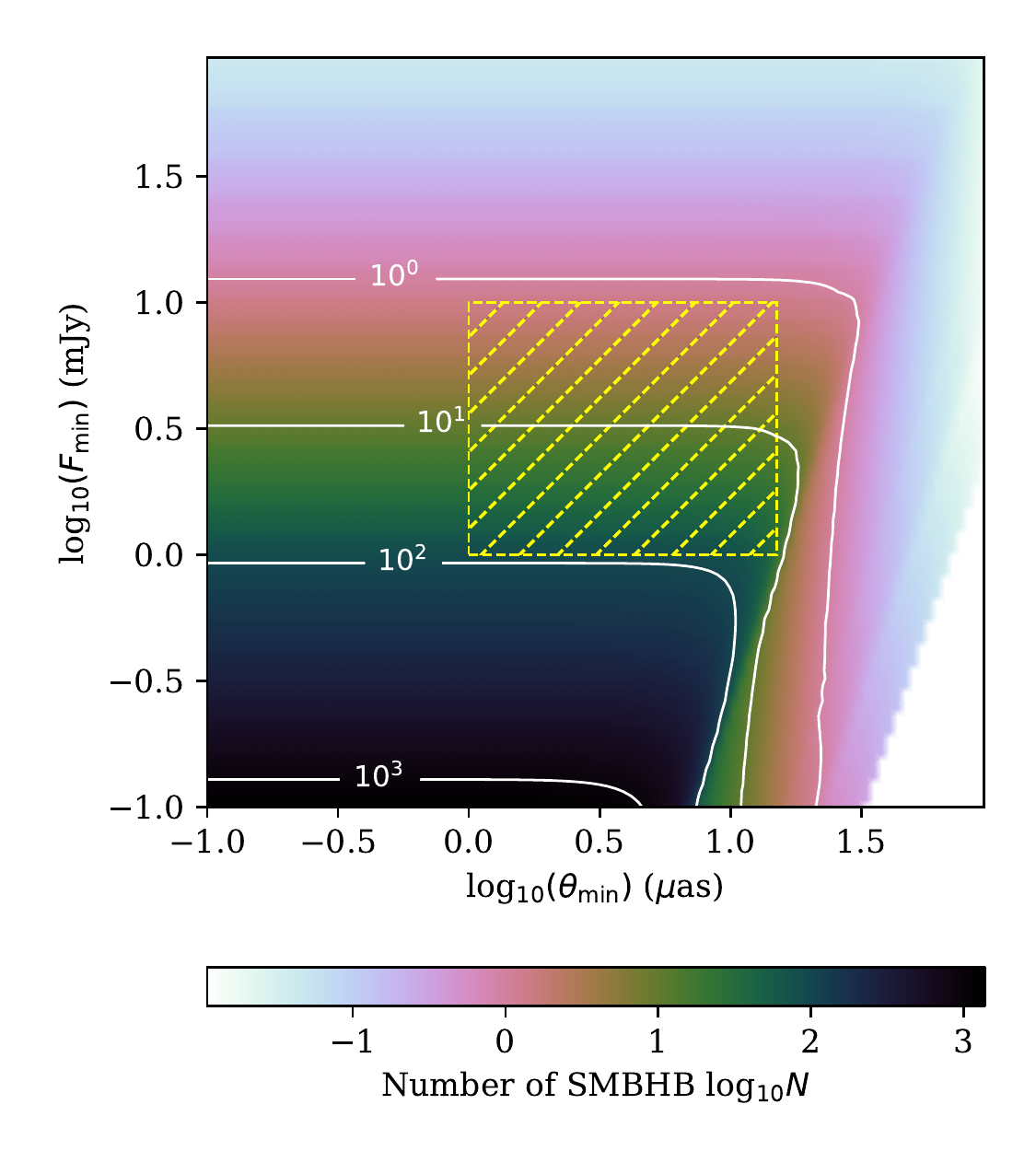}
\caption{Number of detectable SMBHB systems (redshift z<0.5) for the orbital tracking as a function of two main array parameters: the resolution $\theta_{min}$ and the sensitivity $F_{min}$. The hatched area is the target region by ngEHT, where it uses the baseline sensitivity of 10\,mJy and the resolution of 15 microarcsec as the lower limit of the detection number of supermassive black hole binary systems, while the upper limit of the number is roughly corresponding to the array sensitivity and the precision of proper motion measurement by ngEHT, considering a background calibrator in the same field. \label{fig2}}
\end{figure}   
  %\gy{1-sentence comment on maser observations, if possible} 
\begin{itemize}
\item Orbit tracking of supermassive black hole binaries 

The merger of galaxies with central black holes can lead to the formation of a compact supermassive black hole binary (SMBHB) at the new galaxy center \citep{2013ARAA..51..511K}. The early dynamical friction-driven and late gravitational radiation-driven phases of SMBHB evolution are separated by the sub-pc orbital separation regime. How does the SMBHB overcome this regime is known as the final-parsec problem \citep{1980Natur.287..307B}. For ngEHT with SFPR, the propagation delays caused by the troposphere could be canceled out, we can still rely on a signal-to-noise-ratio dependent resolution. Astrometric tracking of a black hole from a SMBHB system can reach 1\,{\textmu}as precision or better \citep{2011ApJ...735...57B, 2018ApJ...863..185D}. In the calculation of a population of detectable SMBHBs, we adopt the fiducial parameters of the model with a larger maximum observed binary period $P_{base}=$30\,yr (see Table 1 in \citep{2018ApJ...863..185D}), and plot the number of SMBHBs as a function of the resolution $\theta_{min}$ and the sensitivity $F_{min}$ (Fig. 2). The ngEHT would provide an opportunity to track several observable sub-pc SMBHBs with a threshold of $\theta_{min}$=15\,{\textmu}as and $F_{min}$=10\,mJy. While considering to track orbit motions of SMBHB with respect to a background source in the same field as the upper limit, the minimum threshold of $\theta_{min}$ and $F_{min}$ is 1\,{\textmu}as (the static ionospheric residuals could be minimized in the in-beam scenario) and 1\,mJy, respectively, as shown in Figure 2.  
%\item Survey of jet apex to black hole distance
%\rd{survey of SFPR astrometry between 85GHz foot of jet to 340GHz BH ring (resolved or otherwise).}
%\wj{Can we also merge this to section 2.3, as an item of $\mu$as relative astrometry?}
%\rd{sure}
\item Relative and absolute astrometric measurements

The direct astrometric output of SFPR is the core-shift. It can be the relative positions between the 85\,GHz core and the photon ring of the black hole when the 340\,GHz already reaches the horizon scale. Otherwise, the core-shift can be used to estimate the magnetic field and the particle density of the innermost jet \citep{Zamaninasab2014}, as well as predict the jet apex up to the infinite frequency \citep{1998AA...330..79L}. This provides a capability to position the black hole and track its motions by synergy with the lower-frequency VLBI, where the absolute astrometry is possible. Meanwhile, the absolute astrometry at short wavelengths needs cluster/paired antennas in each site \citep{2011ApJ...735...57B}. Current proper motions of SgrA* still surffer from the scattering as measured at 43\,GHz \citep{2004ApJ...616..872R,2022arXiv:2210.03390}, if one can go to a higher frequency, this effect can be largely reduced at the ngEHT frequencies. This is also very important to understand the head–tail sources (e.g. IC\,310, NGC\,1265) whose hosting galaxies are infalling into the cluster at a high speed \citep{2020MNRAS.499.5791G}.
\end{itemize}
%\subsection{Detection of cosmic sources at 1mm}
%{figure about how many sources detectable}
%\rd{plot of flux sensitivity/No. sources, for dish size given FTP}
%\wj{estimation from the radio luminosity function (offset exist), the number of AGNs detected is almost in linear to the sensitivity (both in logarithm). Could we merge this section with section 2.2? as an extension from BH shadows towards more faint cosmic sources at 1\,mm. Maybe Figure 3 can be replaced by mentioning this logarithm relationship instead.}
%\begin{figure}[H]
%\includegraphics[width=14 cm]{fig3.pdf}
%\caption{Number of detectable AGNs as a function of the sensitivity $F_{min}$.\label{fig3}}
%\end{figure} 
%\subsection{Survey of Foot of Jet to BH distance}
%\rd{survey of SFPR astrometry between 85GHz foot of jet to 340GHz BH ring (resolved or otherwise).}
%\wj{Can we also merge this to section 2.3, as an iterm of $\mu$as relative astrometry?}
%\rd{sure}
%\subsection{Survey of.. detected}

%\begin{quote}
%This is an example of a quote.
%\end{quote}

%%%%%%%%%%%%%%%%%%%%%%%%%%%%%%%%%%%%%%%%%%
\section{Requirements}

\subsection{Instrumentation requirement}
%\rd{tri-band\\dish size\\sensitivity\\co-located GPS\\co-located RMS path-length measurement (STI)}
The capability of simultaneous observations at a lower frequency band (85\,GHz or 110\,GHz, 3\,mm) and at one or two higher frequency bands (the 255\,GHz or 220\,GHz, 1.2\,mm and 340\,GHz or 330\,GHz, 0.88\,mm) is required for the frequency-phase transfer. This can be accomplished with a quasi-optics tri-band receiving system \citep{2013PASP..125..539H} or a wide band receiver \citep{2021PASJ...73.1116Y}. In the case of a large interferometry array or co-site antennas working as single VLBI station, the capability of forming sub-arrays corresponding to the lower and the higher observing frequency bands is feasible compared to install trip-band receivers for each antenna. A co-located GPS will give accurate site positions for the geometric model, and the root-mean-square of the tropospheric path length fluctuations should be monitored for the co-site antennae. These have been found to greatly reduce the residual ionospheric, positional and tropospheric contributions. Fuller descriptions of their impact can be found in \citep{2001isra.book.....T}. The planned recording data rate as high as 256\,Gbps would be able to incorporate the multi-band data stream simultaneously since the available bandwidth will be shared across all bands. The baseline sensitivity should be high enough to guarantee the fringe detection and minimise phase errors on a correlated flux of $\sim$10\,mJy source at the lower frequency, as well as to achieve a super/over- resolution power \citep{2012AA...541A.135M,2017ApJ...838....1A}. A detailed technical demand on the instruments is presented by Rioja et al. in the same special issue.

\subsection{Strategy of observation and calibration }
SFPR allows a phase calibrator within 10$^{\circ}$ apart in the sky and a switching cycle of more than 10 minutes \citep{2011AJ....141..114R}. SFPR expects a calibrator of correlated flux higher enough at both the low and the high frequency bands that could be fringed. A higher flux is better as mainly to reduce the thermal noise. Meanwhile, a relative large separation, i.e., 10$^{\circ}$ makes it much less restrictive to find a suitable calibrator even at the high frequencies. The core-shift of the phase calibrator would be incorporated into the final core-shift measurement \citep{2015JKAS...48..277J, 2019JKAS...52...23Z}. A prior core-shift of calibrator or a negligible core-shift at RA or DEC direction would be helpful to extract the true core-shift of the target \citep{2018APJL...853..14J,2021APJL...922..16J}. A synergy with the lower-frequency VLBI networks observing simultaneously can obtain more core-shift measurements to fit the power law scheme and perform the absolute astrometry observations.  
%\gy{Add a subsection about "Synergy with the lower-frequency VLBI networks"?}
%\rd{put in a matching ngVLA science case}

\section{Summary} \label{sec:sum}
 With the aid of simultaneous multi-frequency receiving system and more new stations available \citep{2019JKAS...52...23Z, 2020AARv..28....6R}, ngEHT with SFPR technique will be a very powerful tool to investigate the accretion disk and the jet/outflow connection in Sgr\,A* and M\,87*, or other interesting targets at sub-mm wavelengths. With dramatically increased coherence time and more feasible observational requirements (e.g. long switching cycle time and large angular separation of calibrators), it will help to capture more images of black hole shadows and detect black hole motions in a binary system or a galaxy cluster. 

\acknowledgments{ We thank the referees for their constructive comments and suggestions. This work was supported in part by the National Natural Science Foundation of China (grant Nos. 12173074, 11803071, 11933007), the Key Research Program of Frontier Sciences, CAS (grant Nos. QYZDJ-SSW-SLH057, ZDBS-LY-SLH011), the Shanghai Pilot Program for Basic Research - Chinese Academy of Science, Shanghai Branch (JCYJ-SHFY-2022-013), the Spanish Ministerio de Econom\'{\i}a y Competitividad (grants AYA2016-80889-P, PID2019-108995GB-C21), the Consejer\'{\i}a de Econom\'{\i}a, Conocimiento, Empresas y Universidad of the Junta de Andaluc\'{\i}a (grant P18-FR-1769), the Consejo Superior de Investigaciones Cient\'{\i}ficas (grant 2019AEP112), and the State Agency for Research of the Spanish MCIU through the ``Center of Excellence Severo Ochoa" award to the Instituto de Astrof\'{\i}sica de Andaluc\'{\i}a (SEV-2017-0709).}

\begin{adjustwidth}{-\extralength}{0cm}

\reftitle{References}

\end{adjustwidth}
\end{document}